\documentclass[onecolumn]{revtex4}
\usepackage{amsmath,amssymb,graphics,epsfig,subfigure}
\usepackage{color}
\usepackage[colorlinks,linkcolor=red,anchorcolor=red,citecolor=green]{hyperref}
\usepackage{setspace}
\usepackage{booktabs}
\usepackage{float}
\begin{document}
	\thispagestyle{empty}
\begin{center}
	\title{Thermodynamic phase transition rate for the third-order Lovelock black hole in diverse dimensions}
	
	\author{Yu-Shan Wang, Zhen-Ming Xu\footnote{E-mail: zmxu@nwu.edu.cn}, and Bin Wu
		\vspace{6pt}\\}
	
	\affiliation{$^{1}$School of Physics, Northwest University, Xi'an 710127, China\\
		$^{2}$Shaanxi Key Laboratory for Theoretical Physics Frontiers, Xi'an 710127, China\\
		$^{3}$Peng Huanwu Center for Fundamental Theory, Xi'an 710127, China}
	
	\begin{abstract}
	The phase transition has always been a major focus in the study of black hole thermodynamics. This study employs the Kramer escape rate from stochastic processes to investigate the first-order phase transition strength between the large and small black hole states. The results indicate that the phase transition of the third-order Lovelock black holes exhibits significant asymmetric characteristics in diverse dimensions both in the hyperbolic and spherical topology, with an overall trend of  the transition from large black holes to small black holes. Especially for the spherical topology, when the dimension is higher than seven, there exists a certain temperature beyond which a dynamic equilibrium is established for the phase transition. This study provides valuable insights into the first-order phase transition rate of black holes and enriches the understanding of black hole phase transitions.
	\end{abstract}

\maketitle
\end{center}

\section{Introduction}\label{Intro}
Black holes are extreme gravitational entities in the universe, characterized by their intense gravitational fields that prevent even light from escaping. The emergence of the Hawking radiation~\cite{Hawking:1976de}, black hole entropy~\cite{Bekenstein1973ur}, and laws of black hole thermodynamics has provided a deeper understanding of the nature of black holes and has led to intriguing interdisciplinary connections with areas such as thermodynamics and quantum mechanics. In recent decades, significant progress has been made in the study of black hole thermodynamics, with particular attention drawn to phase transitions of black holes in anti-de Sitter (AdS) spacetime. The pioneering Hawking-Page transition has been explored the transition between the large AdS black hole and thermal radiation states~\cite{Hawking1982dh}, while Kastor, et.al. introduced the concept of an extended phase space to investigate van der Waals-like phase transitions in charged AdS black holes~\cite{Kastor2009wy,Dolan2011xt,Kubiznak2012wp}. Furthermore, developments in holographic thermodynamics and constrained phase space thermodynamics within the framework of the AdS/CFT duality have emerged in recent years~\cite{Visser2022,Ahmed2023dnh,Cong2021fnf,Gong2023ywu,Gao2021xtt,Kong2022gwu,Zeyuan2021uol,Kong2022tgt}. These studies have contributed to the construction of a more comprehensive theoretical framework for black hole thermodynamics, offering important new perspectives and innovative outcomes for addressing challenges in this field.

By drawing analogies to classical thermodynamics, black hole thermodynamics posits that (in theory) black holes possess certain microscopic structures. Numerous works have elaborated on this proposition, with Ruppeiner geometry revealing the potential microscopic structure of the AdS black hole and proposing the existence of microstructures termed {\em black hole molecules} in thermal AdS black holes~\cite{Wei2015iwa}. Calculations of the Ruppeiner scalar curvature imply weak mutual attraction between two black hole molecules~\cite{Wei2019uqg,Wei:2019ctz}. Recently, the topology has emerged as a new approach to describing the microscopic structure of black hole phase transitions, utilizing $\Phi$-map topological flow theory to construct topological invariants independent of intrinsic parameters within black holes for topological classification of the same class of black holes~\cite{Wei2021vdx,Wei2022dzw,Chen:2023elp,Yerra2022alz,Yerra2022coh,Wu2023sue,Fang2022rsb,Bai2022klw}. These investigations deepen the understanding of the microscopic structure of black holes and aid in the search for clues to reveal the essence of black holes and quantum gravity theories.

Certainly, we also aim to provide useful clues for the microscopic structure of black holes in the context of black hole phase transitions. Current research on black hole phase transitions is primarily based on the analysis of phase transitions and criticality in classical thermodynamics, while overlooking the detailed description of the phase transition process. Recently the dynamic process of black hole thermodynamics
phase transition has also been preliminarily studied~\cite{Li2020a,Li2020b,Wei2021,Cai2021,Yang2022}. In this study, we investigate the phase transition rates of black holes by addressing the question of which transition probability is greater: from large black holes to small black holes or from small black holes to large black holes. We focus on the third-order Lovelock black hole in diverse dimensions. The Lovelock gravity is an extension of Einstein's gravity to higher-dimensional spacetime, and the third-order Lovelock gravity exhibits a more complex phase transition process~\cite{Myers1988ze,Deruelle1989fj,Cai2003kt,Kastor2010gq,Kastor2011qp,Zou2010yr, Xu2014tja}. Especially in the case of hyperbolic topology in spacetime, the black hole will exhibit a van der Waals-type phase transition. Therefore, in this paper, we take the third-order Lovelock black hole as an example to investigate the rate of phase transitions between the large and small black holes. Hence, before discussing the phase transition rate, it is necessary to introduce the research strategy we have adopted. This leads to an introduction to the generalized free energy~\cite{Xu2021usl,Xu2023vyj} and Kramer escape rate~\cite{Risken,R. Zwanzig} of black holes.

\subsection{Generalized free energy}
\begin{figure}[htb]
	\begin{center}
		\includegraphics[width=80 mm]{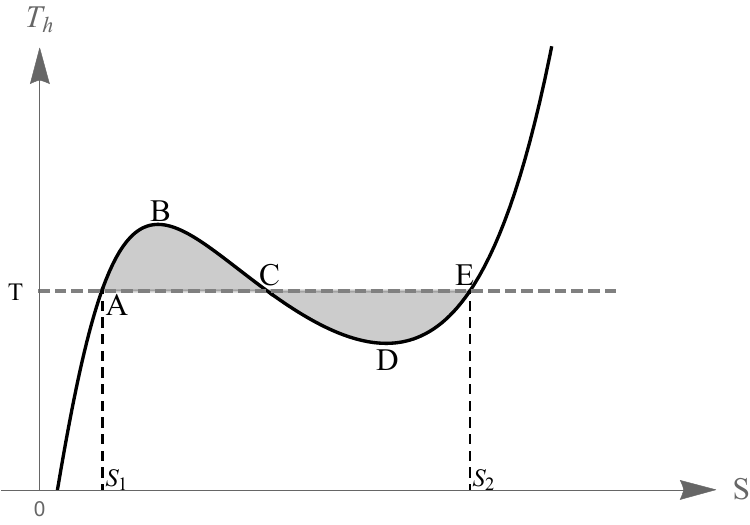}
	\end{center}
	\caption{The Maxwell equal area law of the black hole at the constant pressure.}
	\label{isobaric}
\end{figure}

For a black hole thermodynamic system, the schematic diagram of the equation of state is shown in Fig.~\ref{isobaric}, where the relationship between the Hawking temperature $T_h$ and entropy $S$ of a black hole can be regarded as its equation of state. The Maxwell equal area law states that there exists an isotherm line $T$ such that the $\text{Area}_{\text{ABCA}}=\text{Area}_{\text{CDEC}}$.
This induces an expression
\begin{align}
\int_{S_1}^{S_2}T_h dS=T(S_2-S_1) ~~\Rightarrow~~\int_{S_1}^{S_2}(T_h-T)dS=0
\label{1}
\end{align}

Now we release the upper and lower limits of the above integration and introduce the generalized free energy $U$, which is defined as follows~\cite{Xu2021usl,Xu2023vyj}
\begin{align}
	U=\int(T_h-T)dS,
	\label{U}
\end{align}
where the Hawking temperature $T_h$ of a black hole can be written as a function of entropy $S$, while $T$ is a free parameter. In our previous work~\cite{Xu2021usl}, $T$ is regarded as ensemble temperature of the black hole system and the generalized free energy $U$ is also called as thermal potential. For Eq.~\eqref{U}, we can understand that for a canonical ensemble with temperature $T$, it can be composed of various states, among which when the ensemble temperature equals the black hole Hawking temperature, these states can be regarded as the genuine black hole states, while other states (i.e., $T\neq T_h$) are not. The thermal potential resembles fluctuations in thermodynamics and indicates that all other possible thermodynamic states of the system deviate from the black hole state, and its extremum represents all possible black hole states,
\begin{equation}
	\dfrac{dU}{dS}=0~~\Rightarrow~~T=T_h.
\end{equation}

\begin{figure}[htb]
	\begin{center}
	\subfigure[$T_h-S$ curve]{\includegraphics[width=80 mm]{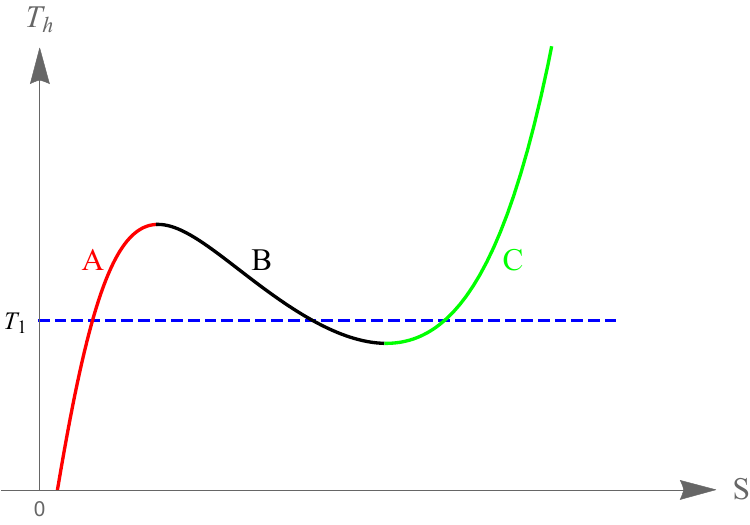}}
\subfigure[The thermal potential curve at $T=T_1$]{	\includegraphics[width=80 mm]{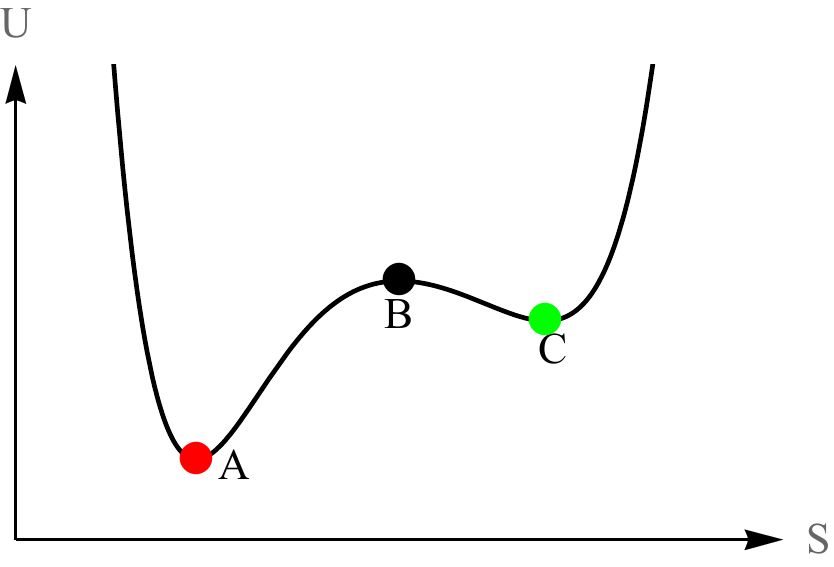}}
	\end{center}
	\caption{The schematic diagram of the thermal potential or generalized free energy $U$ of the black hole.}
	\label{first}
\end{figure}

The schematic diagram of the thermal potential or generalized free energy $U$ described by Eq.~\eqref{U} is illustrated in Fig.~\ref{first}. The red point $A$ on the potential curve corresponds to the red line $A$ on the $T_h-S$ curve, representing a small black hole; The black point $B$ on the potential curve corresponds to the black line $B$ on the $T_h-S$ curve, representing a unstable black hole; The green point $C$ on the potential curve corresponds to the green line $C$ on the $T_h-S$ curve, representing a large black hole~\cite{Xu2023vyj}. The lowest point in the potential curve represents a stable state, the secondary minimum point represents a metastable state, and the local maximum point represents an unstable state. As the parameters of the black hole change, the state of the extremum of the generalized free energy changes, corresponding to the transition between the black hole state and other unknown states in the ensemble.

\subsection{Kramers escape rate}

According to the black hole molecule hypothesis, the phase transition of a black hole is attributed to the rearrangement of black hole molecules caused by thermal fluctuations in the thermal potential. The diffusion motion of black hole molecules is explicitly explained by the Kramers escape rate~\cite{Risken,R. Zwanzig}. As shown in Fig.~\ref{first}, assuming that the ensemble temperature is much lower than the height of the potential barrier, the probability for a molecule to be at the minimum of the potential well far exceeds that of reaching the top of the barrier~\cite{Xu2022jyp}. Even if the molecule reaches the top, it will uniformly fall on either side. If it falls at point $C$, it will stay there for some time before crossing the barrier back to the lowest point $A$. Conversely, it returns from $A$ to $C$.

The probability of transitioning from state $A$ to state $C$ or from state $C$ to state $A$----- the Kramers escape rate, is expressed as
\begin{equation}
	r_k=\frac{\sqrt{\left|U^{\prime \prime}\left(x_{\min }\right) U^{\prime \prime}\left(x_{\max }\right)\right|}}{2 \pi} e^{-\frac{U\left(x_{\max }\right)-U\left(x_{\min }\right)}{D}},
\end{equation}
where $D$ represents the diffusion coefficient, $x_{\min}(x_{\max})$ corresponds to the coordinate position of the minimum (maximum) of the thermal potential, and $U''$ denotes the second derivative of the thermal potential.
Therefore, the probabilities of $A \rightarrow C$ and $C \rightarrow  A$ transitions are expressed as

\begin{equation}
r_{k 1}=\frac{\sqrt{\left|U^{\prime \prime}\left(x_1\right) U^{\prime \prime}\left(x_2\right)\right|}}{2 \pi} e^{-\frac{U\left(x_2\right)-U\left(x_1\right)}{D}}, \quad
r_{k 2}=\frac{\sqrt{\left|U^{\prime \prime}\left(x_3\right) U^{\prime \prime}\left(x_2\right)\right|}}{2 \pi} e^{-\frac{U\left(x_2\right)-U\left(x_3\right)}{D}}
\label{r_k}
\end{equation}
where $x_1$, $x_2$, and $x_3$ represent the positions of $A$, $B$, and $C$, respectively in diagram (b) in Fig.~\ref{first}. In addition $\Delta r_k= r_{k 2}-r_{k 1}$ represents the difference between the rates of two phase transitions, which can help elucidate whether it is  $r_{k 2}$ or $ r_{k 1}$ that governs the black hole phase transition. Without loss of generality, for the numerical calculation of phase transition rate, $D=10$ will be set for all subsequent calculations.

Due to thermodynamic fluctuations, black holes in the thermal potential exhibit some stochastic behavior, reflecting certain thermodynamic characteristics of the black hole. Of course, for the stochastic behavior, we utilize Kramer escape rates described in the Fokker-Planck equation, which characterizes particle Brownian motion in an external field, to study the features of the first-order phase transition rates of black holes~\cite{Xu2022jyp,Liu2023,Du:2023heb}. The structure of this paper is as follows. In Sec.~\ref{Rate}, we discuss the phase transition rates of third-order Lovelock black holes in hyperbolic and spherical topologies. Lastly, in Sec.~\ref{Summary}, we provides a summary and outlook for the entire paper.

\section{Rate of the third-order Lovelock black hole in different topology }~\label{Rate}
At the outset, we briefly review the relevant knowledge of the third-order Lovelock black holes~\cite{Myers1988ze,Deruelle1989fj,Cai2003kt,Kastor2010gq,Kastor2011qp,Zou2010yr, Xu2014tja}, where the action contains the cosmological constant ($\Lambda$), the Einstein-Hilbert term ($R$), the Gauss-Bonnet term ($\mathcal L_2$), and the third-order Lovelock term ($\mathcal L_3$),
\begin{align}
	\mathcal{I} =&\dfrac{1}{16\pi G}\int d^dx(R-2\Lambda+\hat\alpha_2\mathcal L_2+\hat\alpha_3\mathcal L_3),
\end{align}
with the Gauss-Bonnet coupling coefficient $\hat\alpha_{2}$ and the third order Lovelock coupling coefficient $\hat\alpha_{3}$, which can be written as the expressions of the single parameter $\alpha $,
\begin{align}
	\hat\alpha_{2}=\frac{\alpha}{(d-3)(d-4)},\quad
	\hat\alpha_{3}=\frac{\alpha^2}{72{d-3\choose 4}}.
\end{align}
The static spherical symmetry metric for $d \ge 7$ can be read
\begin{align}
ds^2=&-V(r)dt^2+\dfrac{1}{V(r)}dr^2+r^2d\Omega_k^2,\\
V(r)=&k+\dfrac{r^2}{\alpha}\Bigl[1-\Bigl(1+\dfrac{6\Lambda \alpha}{(d-1)(d-2)}+\dfrac{3\alpha m}{r^{d-1}}\Bigr)^{\frac{1}{3}}\Bigr],
\end{align}
where $m$ is a parameter related to the mass of a black hole, and $k=-1$, $0$, and $1$ are topology of the spacetime curvature. The temperature and entropy of the third order Lovelock black hole are~\cite{Wang:2023qxw}
\begin{align}\label{core1}
T_h=&\dfrac{1}{12\pi r_h(r_h^2+k\alpha)^2}\left[\dfrac{48\pi P r_h^6}{(d-2)}+3(d-3)kr_h^4+3(d-5)\alpha k^2r_h^2+(d-7)\alpha ^2k\right] ,\\	\label{core2} S=&\dfrac{\sum_kr_h^{d-2}}{4}\left[1+\dfrac{2(d-2)k\alpha}{(d-4)r_h^2}+\dfrac{(d-2)k^2\alpha^2}{(d-6)r_h^4}\right].
\end{align}
where $P$ is pressure via $P=-\Lambda/(8\pi G)$, and  $\sum_k$ is the volume of the $(d-2)$-dimensional submanifold.

With these preparations in place, we can now proceed to analyze the phase transition rate. According to existing knowledge, for flat topological cases $k=0$, the allowed black hole solutions do not have phase transitions. Therefore, we will focus on two situations: hyperbolic topology and spherical topology.

\subsection{hyperbolic topology}
For the hyperbolic case, existing research suggests that the third-order Lovelock black holes can undergo both first and second-order phase transitions. Therefore, the phase transition processes in dimensions $7<d<12$ are similar. Here, we take $d=7$ as an example for specific investigation.

According to Eqs.~\eqref{core1} and~\eqref{core2}, we take $d=7$ and $k=-1$, and then substitute them into Eq.~\eqref{U}. Hence we have the thermal potential of the third-order Lovelock black holes~\cite{Wang:2023qxw}
\begin{equation}
		\begin{split}
			u&=\dfrac{5}{16}(px^6-3x^4+3x^2-1)-\dfrac{3}{8}t(x^5-\dfrac{10}{3}x^3+5x),
			\label{U1}
		\end{split}
\end{equation}
where these dimensionless parameters are
\begin{equation}
		p:=\dfrac{P}{P_c},\qquad t:=\dfrac{T}{T_c},\qquad x:=\dfrac{r_h}{r_c}, \qquad u:=\dfrac{U}{|U_c|},
		\label{dimensionless}
\end{equation}
here the critical values for the black hole at $d = 7$ are
	\begin{equation}
		P_c=\dfrac{5}{8\pi \alpha},\qquad T_c=\dfrac{1}{2\pi \sqrt{\alpha}},\qquad r_c=\sqrt{\alpha}, \qquad U_c=-\dfrac{11\sum_k}{48\pi}\alpha^2.
		\label{cp1}
	\end{equation}
	
\begin{figure}[htb]
	\centering
			\includegraphics[width=80 mm]{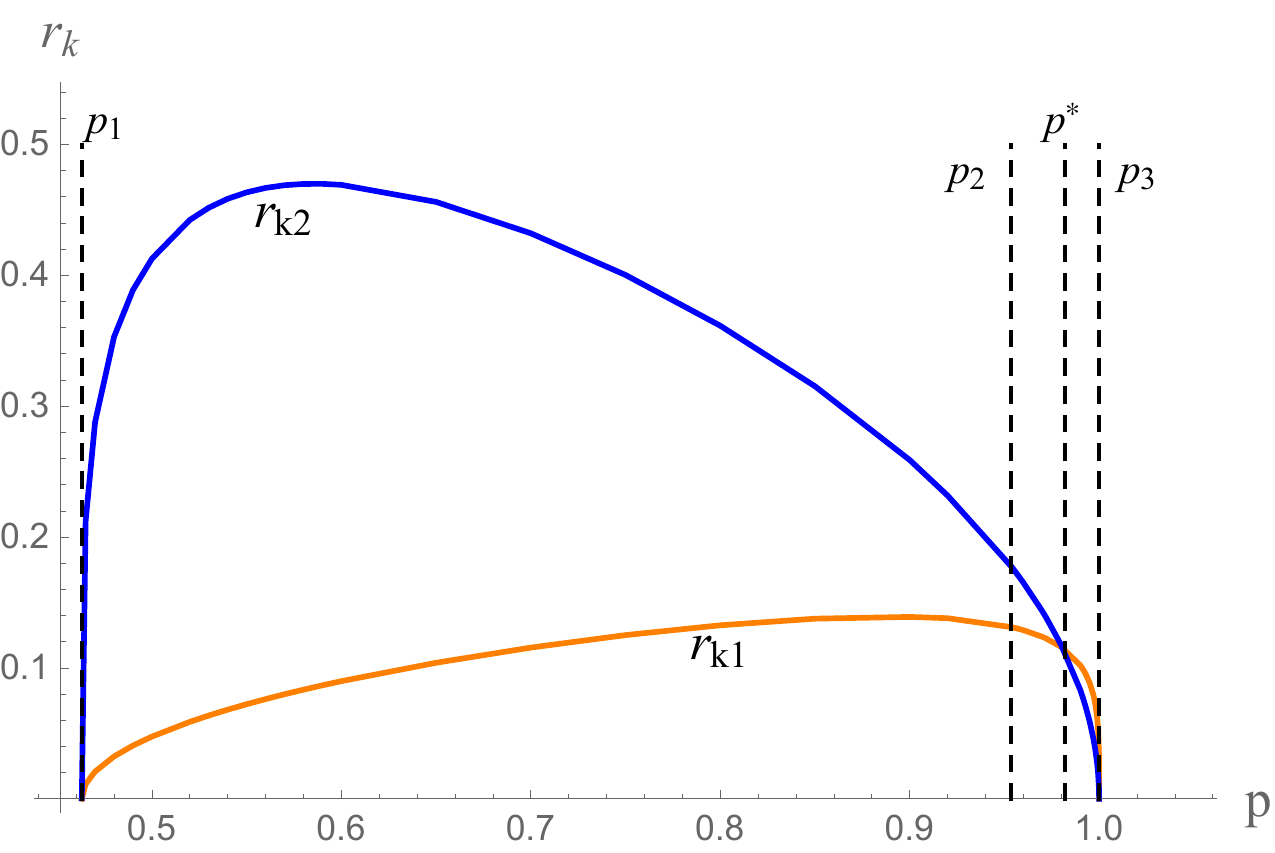}
	\caption{The phase transition rate plots of $t=0.5$ for $d=7$. The solid orange line ($r_{k 1}$) represents the rate of transition from small black holes to large black holes, while the solid blue line ($r_{k 2}$) represents the rate of transition from large black holes to small black holes. }
	\label{-71} 	
	\end{figure}

\begin{figure}[htb]
	\centering
	\subfigure[$p=p_1$]{
		\includegraphics[width=55 mm]{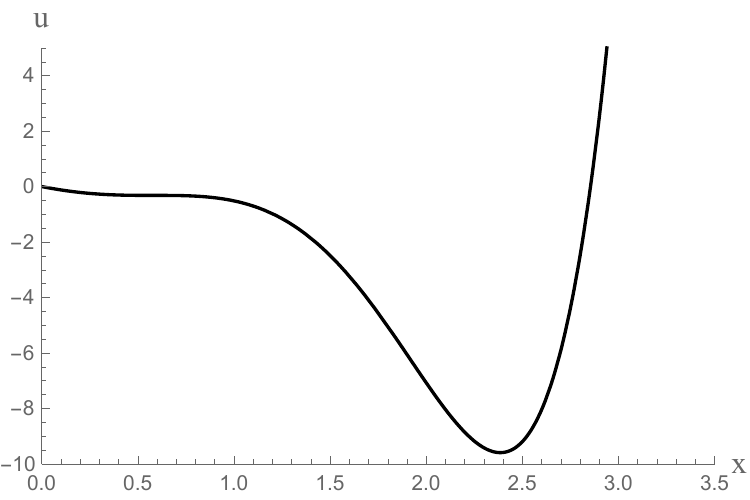} }
	\subfigure[$p=p_2$]{
		\includegraphics[width=55 mm]{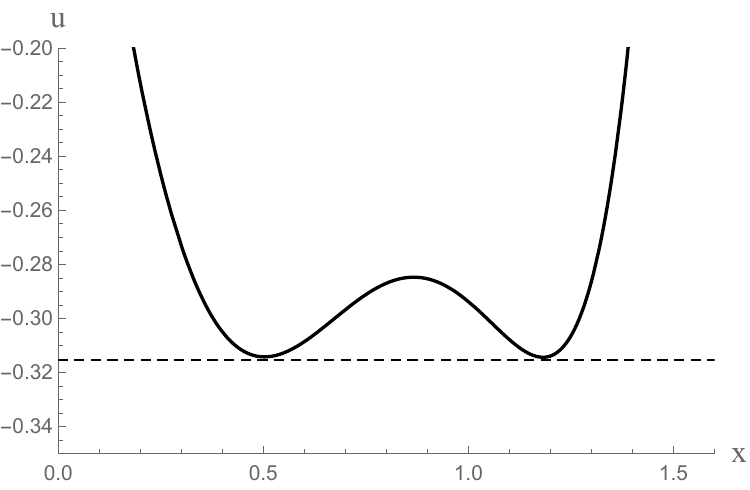}  }
	\subfigure[$p=p_3$]{
		\includegraphics[width=55 mm]{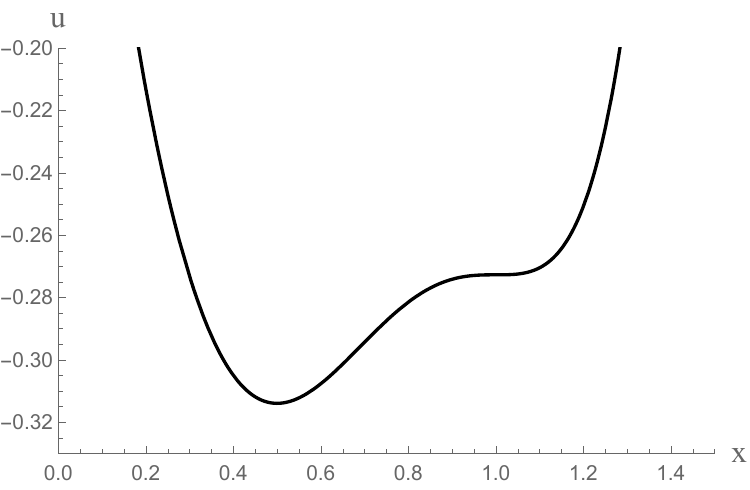}  }
	\caption{The thermal potential of $d=7$ case at $t=0.5$. The pressure $p$ increases gradually from left to right.}\label{-72}	
\end{figure}

After substituting Eq.~\eqref{U1} into the phase transition rate Eq.~\eqref{r_k}, we obtain the phase transition rate of the third-order Lovelock black holes in hyperbolic topology for $d=7$, shown in Fig.~\ref{-71}, where the $p_1$, $p_2$, and $p_3$ represent the three significant nodes of black hole phase transition, and their thermodynamic details are presented in Fig~\ref{-72}.

The phase transition rate $r_{k}$ is zero when the pressure equals $p_1$ and $p_3$, indicating that no phase transition occurs at this point. As the pressure increases from $p_1$ to $p_3$, both rates show an initial increase followed by a decrease. It is observed from the Fig.~\ref{-72} that when $p=p_2$, the system is in a two-phase coexistence state of large and small black holes. While for the Fig.~\ref{-71}, when $p=p_2$, their corresponding phase transition rates are not equal ($r_{k 1} < r_{k 2}$). This indicates that although the thermodynamic system is in a coexisting state of large and small black holes, the overall trend is unstable, with the small black holes about to replace the large black holes as the dominant state of the ensemble. At $p=p^*$, we have $r_{k 1} = r_{k 2}$, indicating that the phase transition reaches dynamic equilibrium at that moment.
	
These phenomena indicate that the system's phase transition exhibits a highly asymmetric characteristic, dominated by the transition from the initially stable state of large black holes to the state of small black holes.

\subsection{spherical topology}
For spherical topology, we have known that the phase transitions in $d=7$ dimensions and those in dimensions greater than 7 should not be conflated. The phase transition at $d=7$ dimensions shares the same type as that with hyperbolic topology, while the phase transitions at $7<d<12$ dimensions belong to the same category, and no phase transition occurs at $d=12$. Next, we will explore the phase transition rate of black holes using the cases of $d=7$ and $d=9$ dimensions as examples.
\subsubsection{7-dimensional scenario}
For the spherical case at $d=7$, the third-order Lovelock black holes can undergo both first and second-order phase transitions. Similar to the case of hyperbolic topology, we obtain the dimensionless thermal potential~\cite{Wang:2023qxw}
\begin{equation}
	\begin{split}
		u&=\dfrac{1}{4}(17px^6+75x^4+15x^2+1)-t(15x^5+10x^3+3x).
	\end{split}
	\label{U2}
\end{equation}

\begin{figure}[htbp]
	\centering
	{
		\begin{minipage}[c]{0.4\linewidth}
			\centering
			\subfigure[]{\includegraphics[width=80 mm]{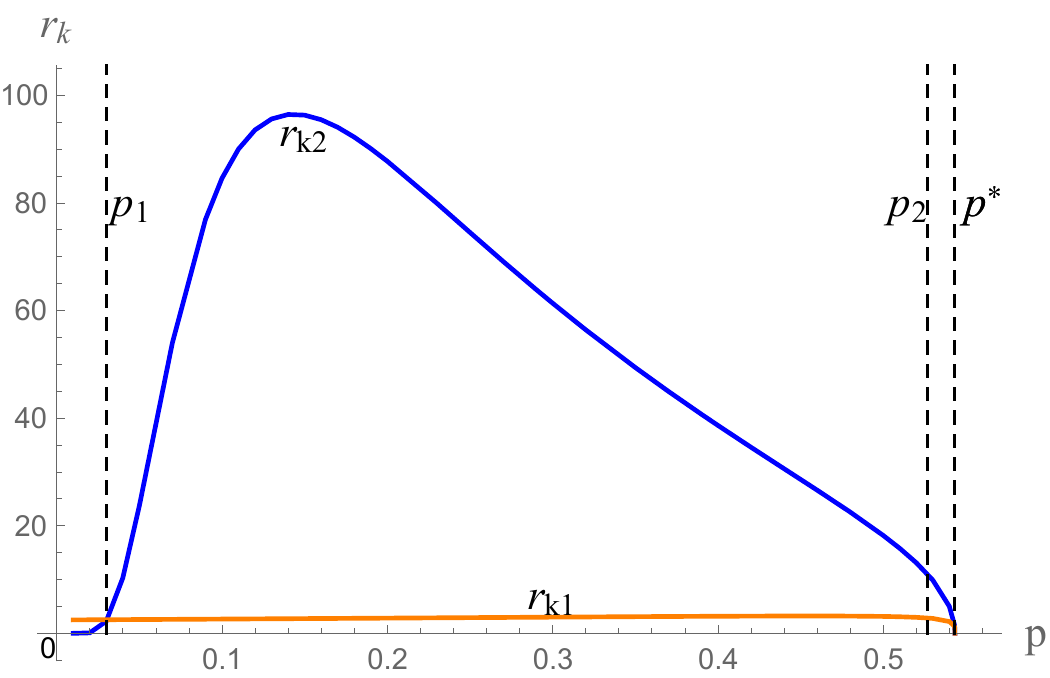}} \label{7a}
		\end{minipage}
		\begin{minipage}[c]{0.4\linewidth}
			\centering	
			\subfigure[]{\includegraphics[width=45 mm]{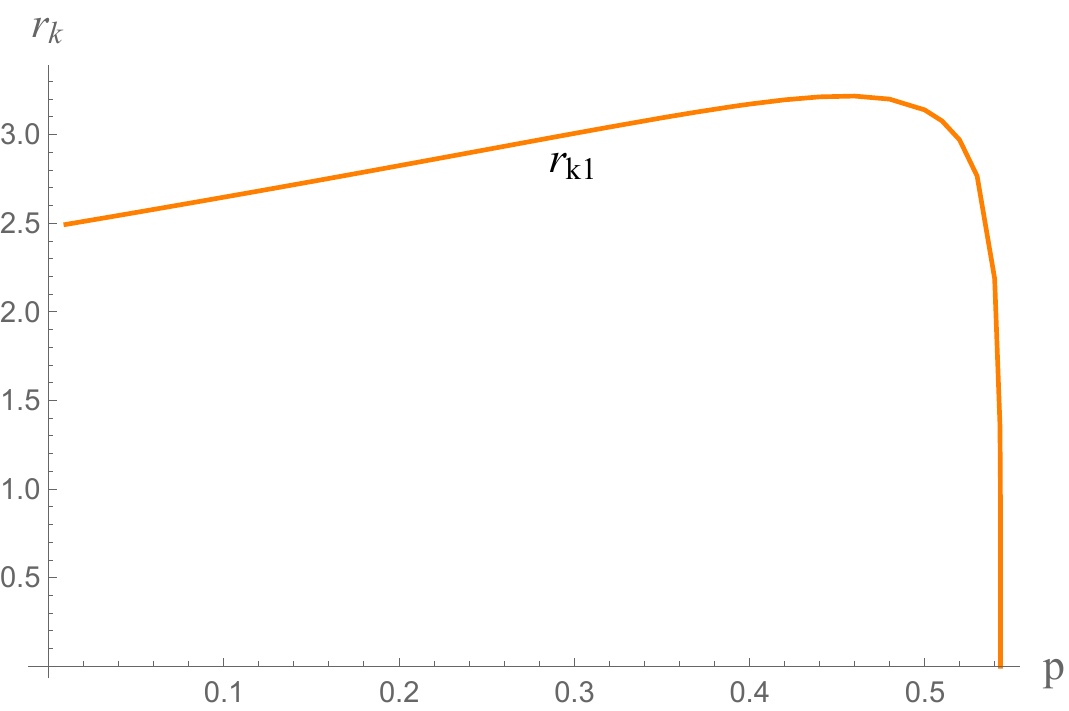}} \label{7b}
			\subfigure[]{\includegraphics[width=45 mm]{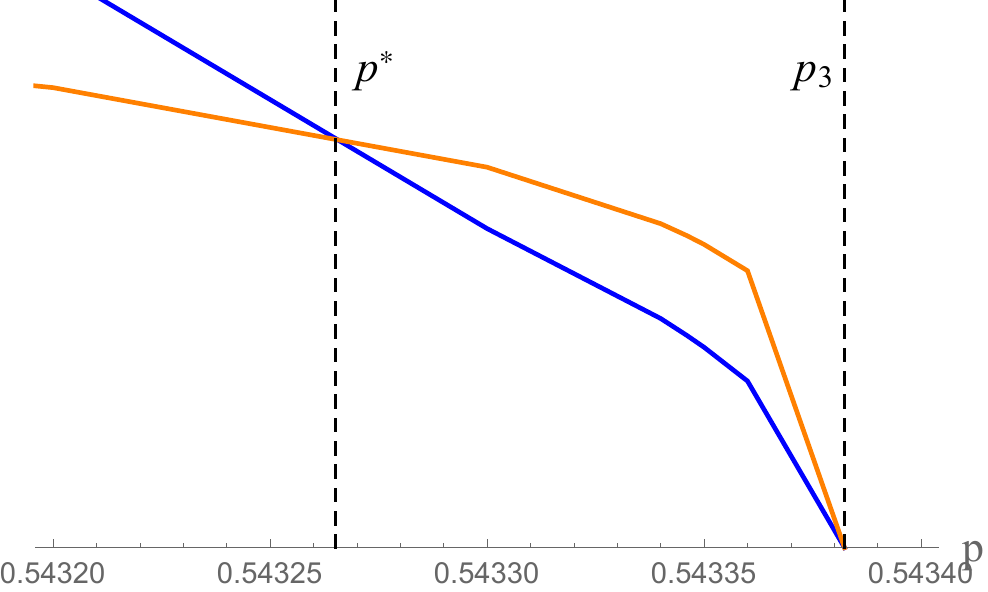} }\label{7c}
		\end{minipage}
	}
	\caption{The phase transition rate diagram at $t=0.8$ for $d=7$. In the diagram (a), the solid orange line ($r_{k1}$) represents the rate of transition from small black holes to large black holes, while the solid blue line ($r_{k2}$) represents the rate of transition from large black holes to small black holes. The diagram (b) is an overall enlarged view of $r_{k1}$. The diagram (c) shows the enlarged details of the intersection at the far right of the the diagram (a).}
	\label{71}
\end{figure}

\begin{figure}[htb]
	\centering
	\subfigure[$p=0$]{
		\includegraphics[width=40 mm]{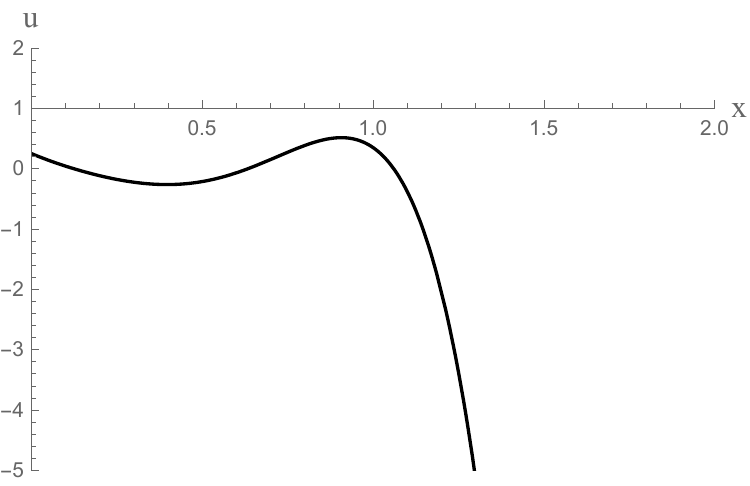} }
	\subfigure[$p=p_1$]{
		\includegraphics[width=40 mm]{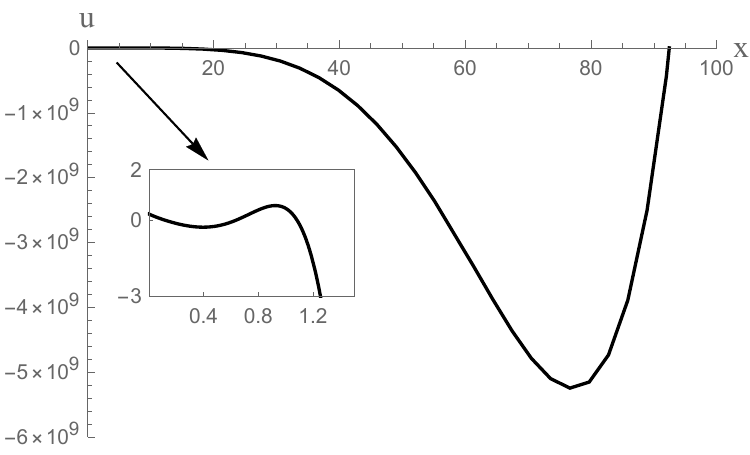} }
	\subfigure[$p=p_2$]{
		\includegraphics[width=40 mm]{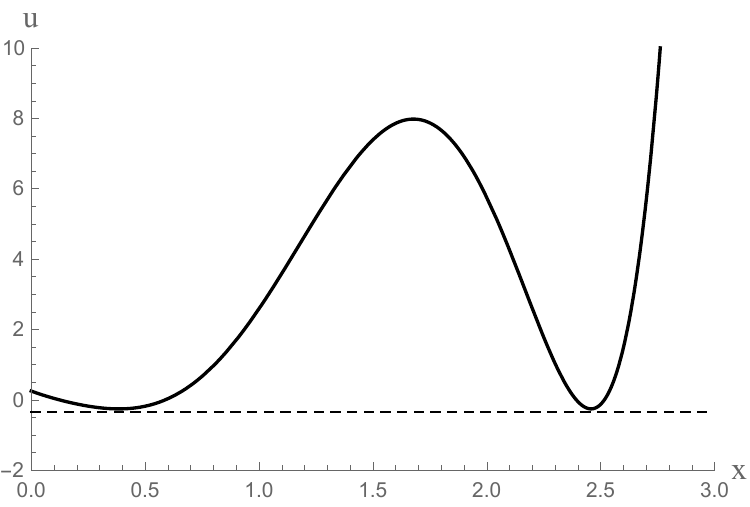}  }
	\subfigure[$p=p_3$]{
		\includegraphics[width=40 mm]{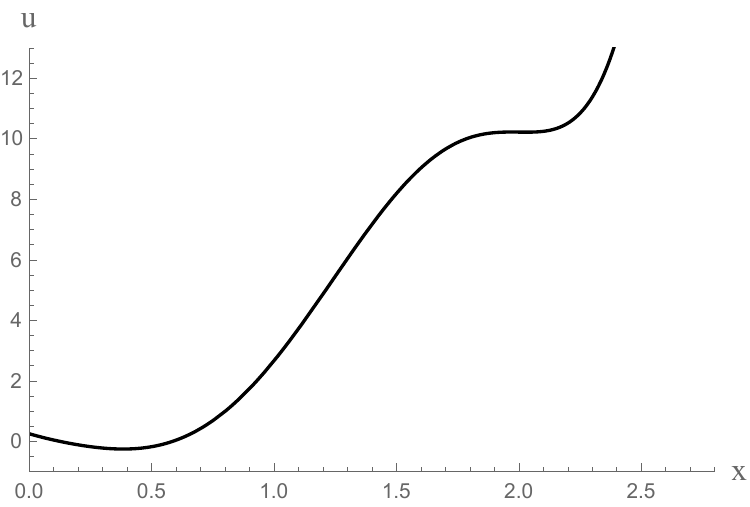}  }
	\caption{The thermal potential of $d=7$ case at $t=0.8$. The pressure $p$ increases gradually from left to right.}
	\label{72}	
\end{figure}

In Fig.~\ref{71}, we present the phase transition rate of the third-order Lovelock black holes in spherical topology for $d=7$. Similar to the hyperbolic case, $p_1$, $p_2$ and $p_3$ represent important nodes of black hole phase transition, as shown in Fig.~\ref{72}. However, unlike the hyperbolic case, the intersection of the two phase transition rate curves at $p_1$ is non-zero. This because that at the initial state with the $p=0$, the thermal potential already contains points corresponding to stable small black holes and metastable black holes, leading to a non-zero transition rate $r_{k1}$. The point corresponding to the large black hole approaches infinitely low values near the initial state, resulting in the transition rate $r_{k2}$ approaching zero. As the pressure increases, the two rates show a trend of first increasing and then decreasing. When $p < p^*$, the corresponding phase transition rates are not equal, with a rate difference  $\Delta r_k=r_{k2}-r_{k1} > 0$. Particularly, when $p = p_2$, although the system is in a coexisting state, the overall trend is unstable, and the probability of the transition from a large black hole to a small black hole is much greater than that from a small black hole to a large black hole. At $p=p^*$, we have $r_{k 1} = r_{k 2}$, and the phase transition reaches dynamic equilibrium at this moment. When the pressure equals $p_3$, the phase transition rate $r_{k} = 0$, and no phase transition occurs at this time.

Similar to the case of hyperbolic topology, the phase transitions exhibit severe asymmetry and are also dominated by the transition process from the original state of a large black hole to a state of a small black hole.

\subsubsection{9-dimensional scenario}
However, for $7<d<12$, we know the phase transition situation can be divided into two types: one is a pure second-order phase transition, and the other is a first-order phase transition accompanied by a second-order phase transition. Different temperatures correspond to different types of phase transitions. Here we choose $d=9$ as the subject, and the thermal potential can be read as~\cite{Wang:2023qxw}
\begin{equation}
	U=\frac{\sum_k}{4}\left[\frac{7}{12\pi}\left(\frac{6}{7}\pi P r_h^8+3r_h^6+3\alpha r_h^4+\alpha^2r_h^2\right)-T\left(r_h^7+\frac{14}{5}\alpha r_h^5+\frac{7}{3}\alpha^2r_h^3\right)\right].
	\label{U3}
\end{equation}
Here, $P$, $T$, and $r_h$ represent the pressure, temperature, and horizon radius of the black hole, respectively.

\begin{figure}[htbp]
	\centering
	\begin{minipage}[c]{0.4\linewidth}
		\centering
		\subfigure[]{\includegraphics[width=80 mm]{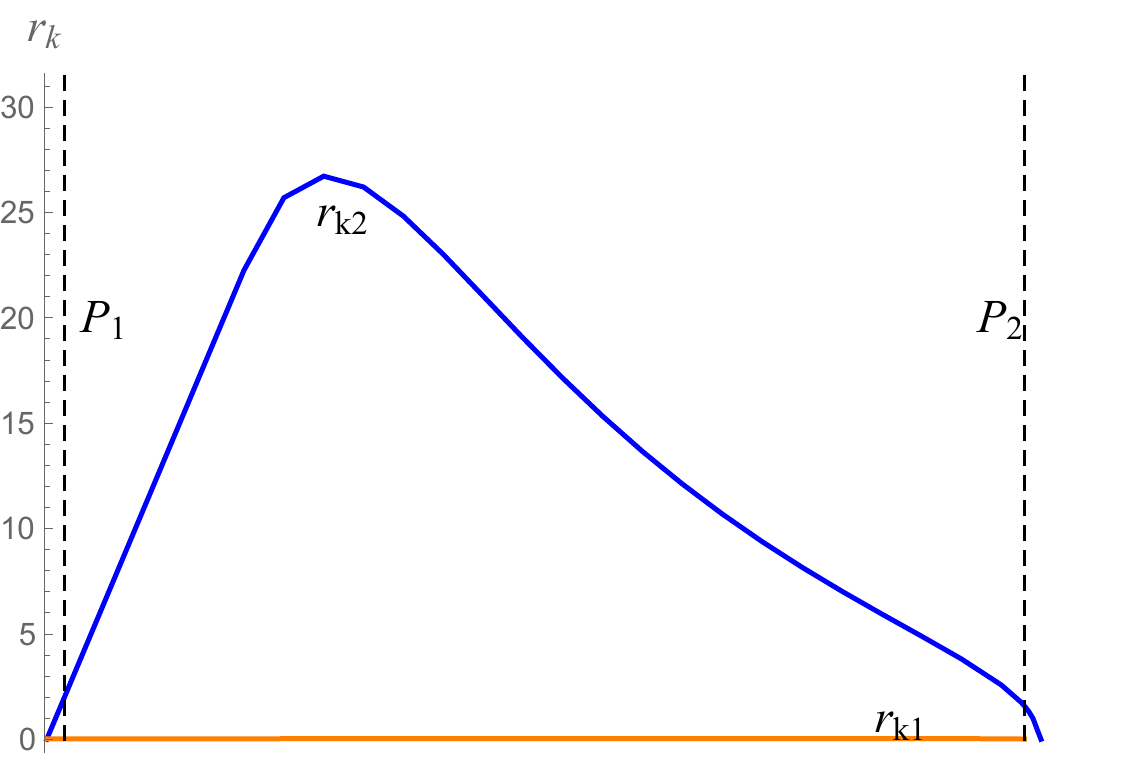} } \label{Tma}
		
	\end{minipage}
	\begin{minipage}[c]{0.4\linewidth}
		\centering	
		\subfigure[]{	\includegraphics[width=45 mm]{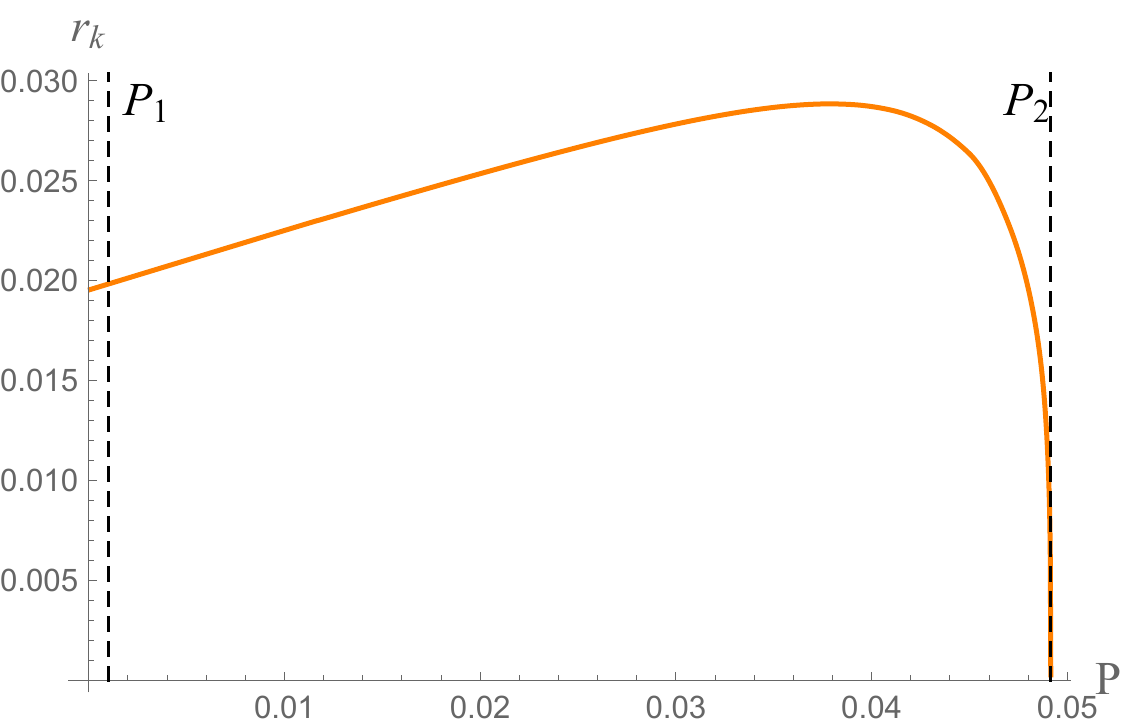}} \label{Tmb}
		\subfigure[]{	\includegraphics[width=45 mm]{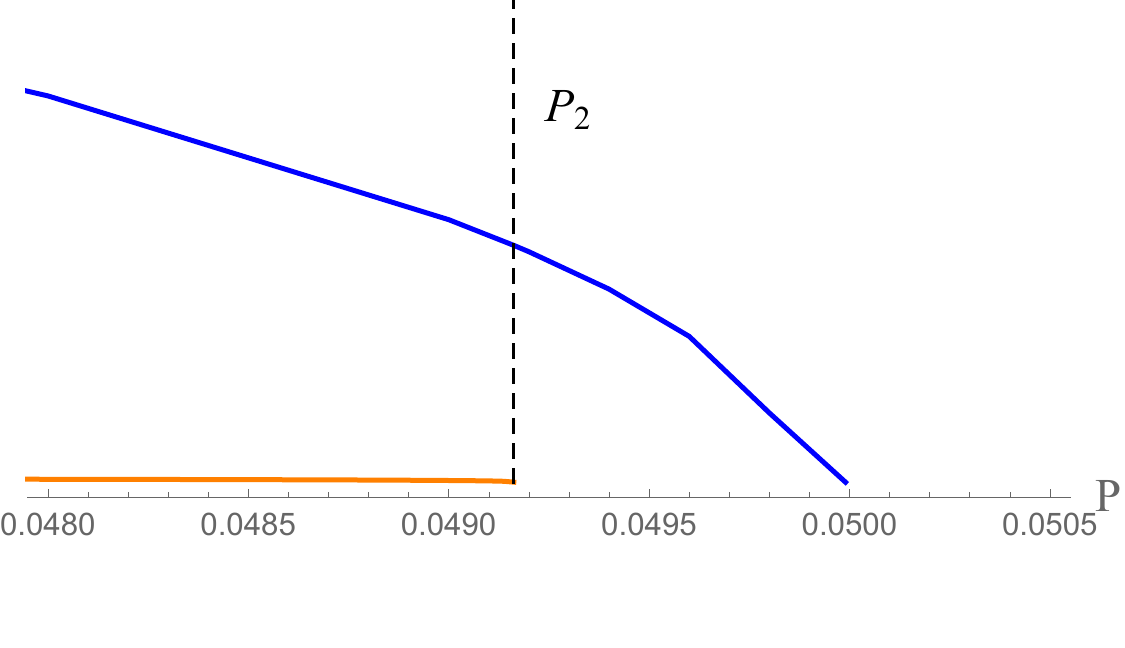} }\label{Tmc}
		
	\end{minipage}
\caption{The phase transition rate diagram at $T=T_m$ for $d=9$ case. In the diagram (a), the solid orange line ($r_{k1}$) represents the rate of transition from small black holes to large black holes, while the solid blue line ($r_{k2}$) represents the rate of transition from large black holes to small black holes. The diagram (b) is an overall enlarged view of $r_{k1}$. The diagram (c) shows the enlarged details of the intersection at the far right of the the diagram (a).}
\label{Tm}
\end{figure}

\begin{figure}[htbp]
	\centering

	\begin{minipage}[c]{0.4\linewidth}
		\centering
		\subfigure[]{	\includegraphics[width=80 mm]{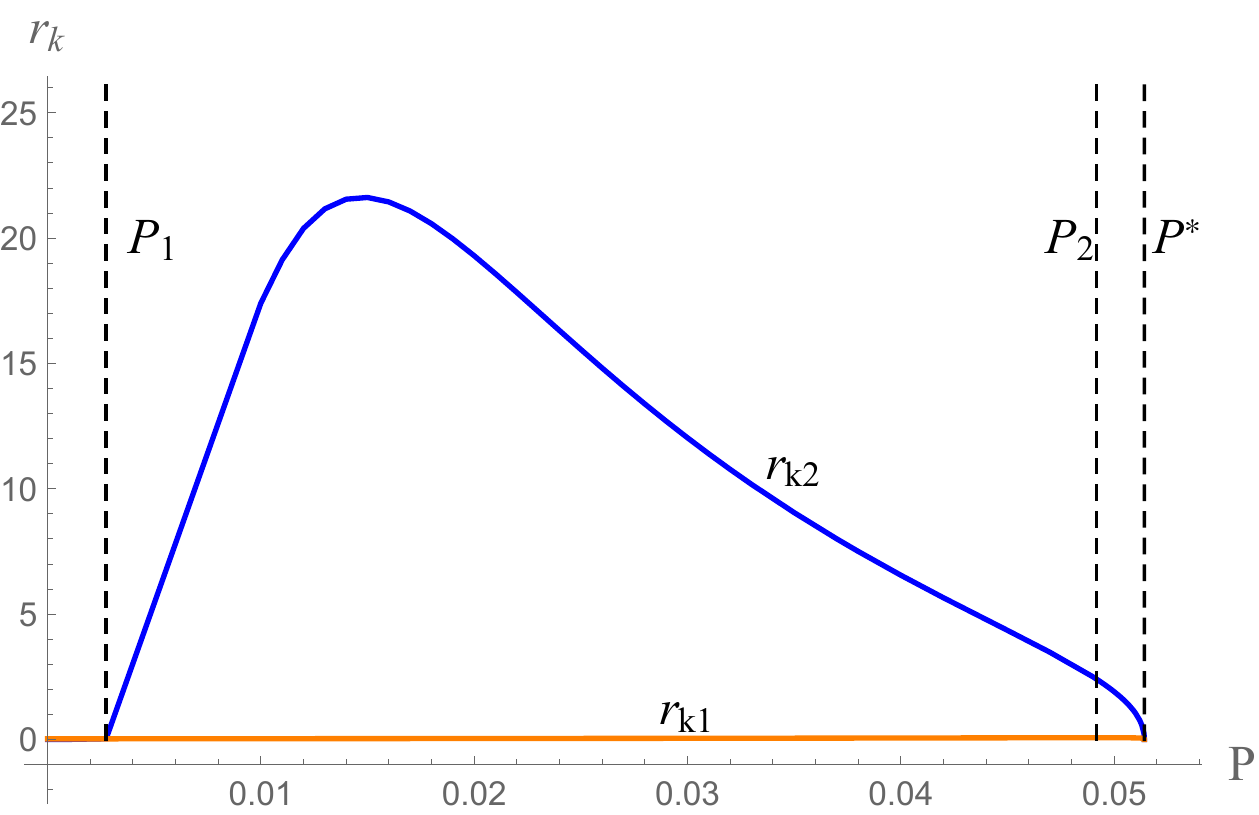} } \label{9a}
		
	\end{minipage}
	\begin{minipage}[c]{0.4\linewidth}
		\centering	
		\subfigure[]{	\includegraphics[width=45 mm]{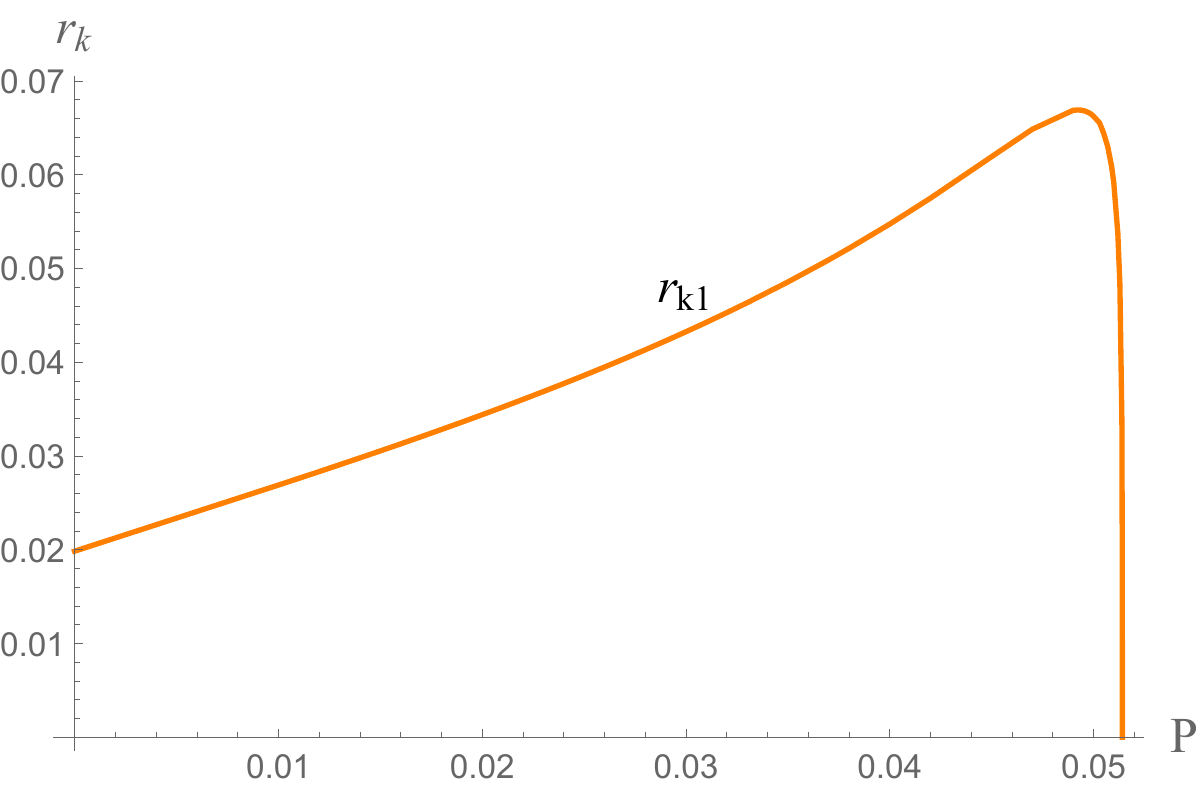}} \label{9b}
		\subfigure[]{	\includegraphics[width=45 mm]{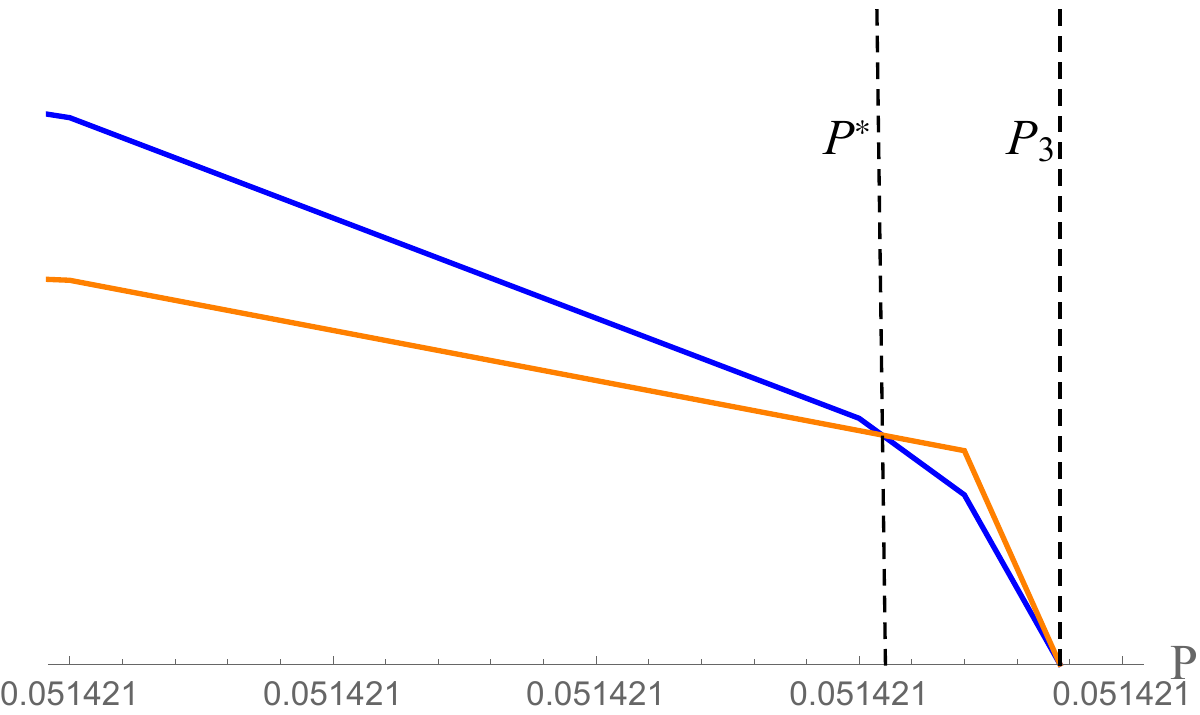} }\label{9c}
		
	\end{minipage}
	\caption{The phase transition rate diagram at $T_m<T=0.21<T_{c2}$ for $d=9$ case. In the diagram (a), the solid orange line ($r_{k1}$) represents the rate of transition from small black holes to large black holes, while the solid blue line ($r_{k2}$) represents the rate of transition from large black holes to small black holes. The diagram (b) is an overall enlarged view of $r_{k1}$. The diagram (c) shows the enlarged details of the intersection at the far right of the the diagram (a).}
	\label{91}
\end{figure}

For spherical topology, when $d>7$, there exist two critical temperatures ($T_{c1}$ and $T_{c2}$), and the phase transition occurs between these two critical temperatures. When $T_{c1}<T<T_m$, it is only the second-order phase transition, while when $T_m<T<T_{c2}$, the system has both second-order and first-order phase transitions. However, in studying phase transition rates, we only consider the first-order phase transition, not second-order phase transition, which simplifies the problem. The phase transition rate diagrams for $T=T_m$ and $T>T_m$ are depicted in the Figs.~\ref{Tm} and~\ref{91}, respectively. In both cases, there exists a non-zero intersection point $P_1$, similar to the 7-dimensional case, which is dependent on the initial state.


When $T=T_m$, in Fig.~\ref{Tm}, we note that as $P$ continuously increases, the phase transition rates $r_{k1}$ and $r_{k2}$ do not intersect. When $P=P_2$, the system is in a coexistence of the large and small black holes, where $r_{k1}=0$ and $r_{k2}>0$, indicating that the system doesn't reach a dynamic equilibrium.

When $T>T_m$, in Fig.~\ref{91}, we observe that the phase transition rate diagram is similar to the Fig.~\ref{71}. For $P_1 < P < P^*$, the corresponding phase transition rates are unequal, with the probability of transitioning from a large black hole to a small black hole being much greater than the reverse process. At $P=P^*$, we have $r_{k1}=r_{k2}$, indicating the phase transition reaches dynamic equilibrium at this time. When the pressure equals $P_3$, the phase transition rate $r_k=0$, and no phase transition occurs.

Evidently, during the transition from Fig.~\ref{Tm} to Fig.~\ref{91}, there exists a certain temperature $T_m$ beyond which a dynamic equilibrium is established for the phase transition.

\section{Summary}~\label{Summary}
We have known the phase transition of large and small black holes in the context of third-order Lovelock black holes, but we know very little about the detailed discussion on the dominant process driving the black hole phase transition. As an extension of the phase transition of black holes, we further explored the phase transition rates of the third-order Lovelock black holes. Based on the black hole molecule hypothesis, we treat the black hole phase transition as a stochastic process of molecular Brownian motion and introduced thermal potential and Kramer rate to study the first-order phase transition rates. Overall, the first-order phase transition rate of the third-order Lovelock black hole shows a trend of increasing first and then decreasing.

By using Eq.~\eqref{r_k}, where $r_{k1}$ represents the probability of a molecule transitioning from $A$ to $C$ and eventually staying at $C$ and $r_{k2}$ represents the probability of a molecule transitioning from $C$ to $A$ and eventually staying at $A$ in Fig.~\ref{first}, the relationship of the first-order phase transition rate between large and small black holes of the third-order Lovelock black hole is obtained.
\begin{itemize}
	\item When the thermodynamic pressure $p< p^*$, we obtain $r_{k2}>r_{k1}$, indicating that the phase transition rate from large black hole to small black hole is much greater than the phase transition rate from small black hole to large black hole.
	
	\item As thermodynamic pressure $p$ increases, the two rates become equal at $p=p^*$, indicating that the phase transition process has reached dynamic equilibrium at this point.
	
	\item When the thermodynamic pressure $p>p^*$, the phase transition from small black hole to large black hole becomes dominant.

    \item For the spherical topology, when $d>7$, there exists a certain temperature $T_m$ beyond which a dynamic equilibrium is established for the phase transition.
\end{itemize}

For third-order Lovelock black holes in diverse dimensions in different topologies, the results reveal significant asymmetric characteristics in phase transition rate, and the first-order phase transition is dominated by the transition process from the initial large black hole state to the small black hole state. This is different from the that of charged AdS black holes, where the transition from small black holes to large black holes is dominant~\cite{Xu2022jyp}. We believe that this research approach is applicable to other black hole models and can introduce new methods into the field of black hole thermodynamic phase transitions, thereby deepening and enriching the content of black hole thermodynamics.

\end{document}